\documentclass[10pt,preprint]{aastex}

\newcommand{\be}{\begin{equation}}
\newcommand{\ee}{\end{equation}}
\newcommand{\ba}{\begin{eqnarray}}
\newcommand{\ea}{\end{eqnarray}}
\newcommand{\bea}{\begin{eqnarray}}
\newcommand{\eea}{\end{eqnarray}}

\slugcomment{To appear in the Astrophysical Journal}

\begin{document}

\title{Nonlinear Couplings Between r-modes of Rotating Neutron Stars}

\author{Sharon M. Morsink}

\affil{Theoretical Physics Institute,
	Department of Physics,
	University of Alberta\\
	Edmonton, AB,
	T6G 2J1, Canada}

\begin{abstract}
The r-modes of neutron stars can be driven unstable by gravitational
radiation. While linear perturbation theory predicts the existence of
this instability, linear theory can't provide any information about
the nonlinear development of the instability. The subject of this paper is the 
weakly nonlinear regime of fluid dynamics. In the weakly nonlinear regime,
the nonlinear fluid equations are approximated by an infinite set of
oscillators which are coupled together so that terms quadratic in the
mode amplitudes are kept in the equations of motion. In this paper, the
coupling coefficients between the r-modes are computed. The stellar
model assumed is a polytropic model where a source of buoyancy is 
included so that the Schwarzschild discriminant is nonzero. The properties
of these coupling coefficients and the types of resonances possible are
discussed in this paper. It is shown that no exact resonance involving
the unstable $l=m=2$ r-mode occur and that only a small number of modes
have a dimensionless coupling constant larger than unity. 
However, an infinite number of resonant mode
triplets exist which couple indirectly to the unstable r-mode.
All couplings in this paper involve the $l>|m|$ r-modes which only
exist if the star is slowly rotating. 
This work is complementary to that of \citet{Sch01}
who consider rapidly rotating stars which are neutral to convection. 
\end{abstract}

\keywords{instabilities --- stars: neutron --- 
stars: oscillations --- stars: rotation}

\section{Introduction}

Gravitational radiation can drive nonaxisymmetric normal
modes of rotating star unstable \citep{Cha70,FS78b,Fri78} (CFS).
The CFS instability affects nonaxisymmetric  modes of stars which  
counter-rotate when viewed in a reference frame rotating with the star, but
co-rotate when viewed in the inertial frame. The nonaxisymmetric
modes are predicted to radiate gravitational radiation and the emission 
of radiation drives the instability. This provides a mechanism for 
converting the star's rotational energy into gravitational radiation, which
will cause the star to spin down to slower angular velocities while providing
a strong source of gravitational radiation which could potentially be 
detected.

The traditional analysis of the CFS instability has been in terms of 
the instability of the fundamental pressure modes (f-modes) of
rapidly rotating neutron stars (eg. Stergioulas 1998). 
However, it was suggested
by \citet{And98} that fluid modes of neutron stars driven by
the Coriolis force (Rossby waves, or r-modes) can also be driven 
unstable by the CFS mechanism. It has since been shown that the
r-modes are indeed unstable within the context of linear 
perturbation theory (Friedman \& Morsink 1998) and that the
growth times are short enough to be of astrophysical interest
(Lindblom, Owen \& Morsink 1998; Andersson, Kokkotas \& Schutz 1999).
A review of recent results on the astrophysical relevance of 
the r-mode instability has been given by Andersson \& Kokkotas (2001).

The use of linear perturbation theory is sufficient to show 
that the modes are unstable, however a linear analysis can not 
describe the evolution of the instability when the amplitude of the
perturbation grows large.  It is inevitable, then, 
that linear perturbation theory
can not be a sufficient description of an unstable system. In order to 
fully describe the growth and saturation of an instability, it is necessary
to incorporate the nonlinear aspects of the system, either through
a full numerical evolution of the nonlinear equations describing the
system, or through a higher order perturbation expansion.

One of the most interesting aspects of the r-mode instability to
gravitational radiation reaction, is the prediction (Owen et. al. 1998) 
that the
gravitational radiation could be detected with an advanced version
of LIGO. However, the analysis made by Owen et. al. (1998) is dependent
on the maximum amplitude the unstable mode is assumed to have before 
nonlinear effects dominate. The value of the maximum mode amplitude is
inaccessible from a purely linear perturbation analysis and it is 
necessary to consider the nonlinearities in the fluid equations
of motion. First steps towards including nonlinearities in the
study of r-modes have been made. Two numerical evolutions of the
fully nonlinear fluid equations have recently been made. 
The first evolution (Stergioulas \& Font 2001) is fully relativistic,
however, changes in the gravitational field induced by the perturbation
had to be neglected in order to make the computation feasible. This
evolution found that over the course of 25 r-mode pulsation periods,
no saturation was observed. A second numerical evolution 
(Lindblom, Tohline and Vallisneri 2001a, 2001b) was performed using Newtonian
gravity and post-Newtonian gravitational radiation reaction terms
and did not show any signs of amplitude saturation due to 
nonlinearities. Neither numerical evolution can definitively explain
the lack of nonlinear saturation.

An alternative approach to the nonlinear evolution problem is to
consider the weakly nonlinear regime and to do a higher order 
perturbation analysis. This approach involves taking the 
general Hamiltonian for a rotating fluid and perturbing around 
the equilibrium. This allows the explicit evaluation of 
coupling coefficients between different modes, so that it is 
possible to make predictions about the maximum amplitudes 
allowed before the perturbation analysis breaks down. This
paper will apply this approach to the nonlinear coupling 
between the r-modes. This paper is a complementary study to that
of \citet{Sch01} who have derived the Hamiltonian
for nonlinear perturbations of rotating Newtonian stars. 
\citet{Sch01} have explicitly evaluated the lowest order 
nonlinear coupling coefficients for stars which have a small
ratio of $N/\Omega$, where $N$ is the Brunt-Vaisala 
frequency (due to buoyancy) and $\Omega$ is the star's 
angular velocity.
In this paper, the opposite limt is considered. This
is equivalent to a slow-rotation approximation, 
while the work of \citet{Sch01} corresponds to 
rapid rotation.
The results of  numerical evolutions of the interactions between 
the coupled modes considered in this paper and those
considered by \citet{Sch01} will be presented
in a following paper (Arras et. al. 2002).

The weakly nonlinear regime has been investigated by other
authors for other types of modes and stars. The role of 
parametric resonances in the nonlinear coupling of modes
of nonrotating stars were discussed in detail 
by \citet{Dzi82}. This work showed how stable equilibrium
solutions can lead to mode saturation and was extended
to describe the behaviour of Cepheids by \citet{DK84}. 
The role of nonlinear interactions between p-modes in
the Sun were investigated by \citet{KG89} and \citet{KGK94}.
Amplitude saturation of g-modes in white dwarfs through
nonlinear interactions was shown in a paper by \citet{WG01}.
All of the above calculations have involved only the lowest
order nonlinear terms. \citet{VH94} has constructed a 
higher order Hamiltonian which includes terms which are
quartic in the displacement. Higher order terms have
also been included in a relativistic study of radial
oscillations \citep{SPA01}.
None of the above studies has considered rotating stars. 
However,
in order to be able to study the saturation of an unstable
r-mode, it is necessary to include the complicating
effect of rotation. In geophysics, r-modes in oceans
are of great interest and nonlinear interactions between
the r-modes in oceans have been studied by 
\citet{LG67}, \citet{Rip81}, and \citet{Pok95}. However, these calculations 
are typically done in the ``$\beta$-plane'' approximation where
locally the surface of a sphere is replaced by a plane. 
This approximation does not describe very well the 
situation of interest in stellar physics, where the 
star as a whole is expected to oscillate. In the
calculations where the curvature of the ocean is 
not neglected, the radial dependence of the mode's eigenfunction
is ignored, and this is not valid for stellar physics. 
As already 
discussed above, \citet{Sch01} are the first to study 
the effect of nonlinear interactions amongst the
modes of rotating stars.

The structure of this paper is as follows. In section \ref{lin}
linear perturbation theory is briefly reviewed. In section
\ref{third} an alternative form of the third order Hamiltonian
is derived. In section \ref{eom} the form of the nonlinear 
equations of motion for oscillations of a rotating star are
reviewed. The type of resonances allowed in the coupled
r-mode system are discussed in section \ref{res}. Numerical
results are presented in section \ref{num} and their 
implications are discussed in the conclusion.

\section{Linear Perturbation Theory of Rigidly Rotating Nonrelativistic Fluids}
\label{lin}

The equilibrium state of a rotating star is described by its 
pressure $p$, density $\rho$, velocity $v^a$ and gravitational field $\Phi$. 
In this work, only rigidly rotating stars are considered, so that the
star's velocity field is of the form $v^a = \Omega \phi^a$ where $\Omega$
is a constant angular velocity. Small perturbations about the equilibrium 
star are described by the displacement vector field $\xi^a$ which connects
a fluid element in the equilibrium star to a fluid element in the perturbed 
star.

The Hamiltonian for the perturbation theory of a nonrelativistic fluid 
is a series expansion in the fluid displacement field of the form
\be
\delta H = \delta H_2(\xi^a,\xi^a) + \delta H_3(\xi^a,\xi^a,\xi^a)
	+\; ...\; + \delta H_n(\xi^a,...,\xi^a) + O(|\xi|^{n+1}),
\ee
where each term in the series depends on the equilibrium fluid variables 
and $\xi$. The subscript on each term denotes the order at which the
displacement field $\xi^a$ appears in the term. If all terms up to 
and including the term $\delta H_n$ are kept in the expansion, then the
resulting equations of motion will be of order $(n-1)$ in  $\xi^a$.
In this notation, retaining only the second order Hamiltonian term 
results in linear perturbation theory. 

The second order Hamiltonian for a nonrelativistic rigidly rotating 
fluid  is \citep{FS78a}
\begin{eqnarray}
\delta H_2 &=& \frac{1}{2} \int \left[
	\rho |\dot\xi |^2 
	+ \frac{1}{\rho} \delta \rho \delta p 
	- \frac{1}{4\pi G} |\nabla \delta \Phi|^2 \right] dV, 
\label{ec1}\end{eqnarray}
where  the 
Eulerian perturbations of the density and pressure are given by
\bea
\delta \rho &=& - \nabla_i (\rho \xi^i)\\
\delta p    &=& - \Gamma_1 p \nabla_i \xi^i - \xi^i \nabla_i p,
\eea
and the perturbed gravitational potential satisfies
\be
\nabla^2 \delta \Phi = 4 \pi G \delta \rho.
\ee
The expression (\ref{ec1}) can be physically interpreted 
as the energy of the perturbation as measured in the rotating frame and
has also been named the rotating frame 
canonical energy by \citet{FS78a}, where its
properties have been discussed extensively.

\section{An Alternative Expression for the Third Order Hamiltonian}
\label{third}

The expansion of the Hamiltonian up to and including terms third order in the 
displacement $\xi$ has been found by Kumar \& Goldreich (1989) for the case of 
static fluids\footnote{The terms involving the perturbed gravitational field
given by \citet{KG89} are incorrect  and have been corrected by \citet{Sch01}.}. 
When the equilibrium star is rotating, the contributions to the perturbed 
Hamiltonian through the perturbation's kinetic energy is altered. 
Surprisingly, \citet{Sch01} have shown that the contribution from the
kinetic energy of the perturbation is exactly second order in the displacement. 
As a result, all terms in the perturbation expansion of the Hamiltonian for
a rotating star which
are third order in the displacement (or higher order) are exactly the same as the
terms in the perturbation expansion for a nonrotating star. This is an important 
result and simplifies the study of perturbations of rotating stars greatly.
The third order term in the perturbation expansion of the Hamiltonian is
\ba
\delta H_3 
	&=& \frac16 \int dV \left[ -2 \; p \; \nabla_a \xi^b \nabla_b \xi^c \nabla_c \xi^a
		+ \rho \;  \xi^a \xi^b \xi^c 
		\nabla_a \nabla_b \nabla_c \Phi \right.  \label{deltaE}\\
	&&  \qquad  -3 \; p \; (\Gamma_1-1) \;  \nabla \cdot \xi \;
		\nabla_a\xi^b \nabla_b\xi^a 
		- \; p \; \left( (\Gamma_1-1)^2 + \frac{\partial\Gamma_1}{\partial \ln \rho}\right) \; 
		(\nabla \cdot \xi)^3  \nonumber \\
	&&   \qquad \left. + 3 \;\rho \; \xi^i \xi^j \nabla_i \nabla_j \delta \Phi\right],\nonumber
\ea
as derived by \citet{KG89} and corrected by \citet{Sch01}.

Although the third order Hamiltonian (\ref{deltaE}) describes the nonlinear
interactions of modes, it is not simple to see from this expression what 
the relative size of the interaction term is to other physical energy scales. 
For this reason, an alternative form of equation (\ref{deltaE}) will now
be presented which is useful for finding the size of the nonlinear
coupling between different r-modes.

The goal is to replace the first term in (\ref{deltaE}) with a simpler 
expression by integrating by parts.
The explicit calculation can be found in the Appendix.
The final result is
\ba
- 2 \int dV  p \; \nabla_a \xi^b \nabla_b \xi^c \nabla_c \xi^a
	&=& \int dV \left[
		\xi^a \xi^b \xi^c \nabla_a \nabla_b \nabla_c p 
		+ 3 (\nabla \cdot \xi) \; \xi^b \xi^c \nabla_c \nabla_b p
	\right. 	\nonumber \\
	&& \qquad	- 3 \nabla \cdot (p\xi) \nabla_a \xi^b \nabla_b \xi^a
		+ 2 (\nabla\cdot\xi)^2\; \xi \cdot \nabla p 	\nonumber \\
	&& \qquad 	\left. - 2 p\;(\nabla \cdot \xi) \; \xi^a \nabla_a (\nabla \cdot \xi)
	\right]. \label{triplep}
\ea
While the expression (\ref{triplep}) doesn't appear to be a simplification, 
it is useful,
since the first term involves three derivatives of the equilibrium pressure. 
Recall that
the third order Hamiltonian (\ref{deltaE}) has a term involving 
three derivatives of the equilibrium gravitational potential. This suggests that these terms
be combined using the equilibrium equations of motion. 
The equation of hydrostatic equilibrium for a rotating
star is 
\be
\frac12 \nabla_a (v^2) + \frac{1}{\rho}\nabla_a p + \nabla_a \Phi = 0.
\label{hydro}
\ee
The strategy will be to take two covariant derivatives of the equation of 
hydrostatic equilibrium and use the result to simplify the 
third order Hamiltonian.

In the case of rigid rotation, 
\be
\nabla_a \nabla_b \nabla_c (v^2) = 0.
	\label{zero}
\ee
To prove equation (\ref{zero}), use Cartesian coordinates, so that 
$v^2 = \Omega^2(x^2+y^2)$ and $\nabla_a = \partial_a$. Then
\ba
\nabla_a(v^2) &=& 2 \Omega^2 ( x \partial_a x + y \partial_a y)\\
\nabla_b \nabla_a(v^2) &=& 2 \Omega^2 (\partial_a x \partial_b x
	+ \partial_a y \partial_b y)
\ea
from which eq. (\ref{zero}) follows.

Taking the two derivatives of the equation of hydrostatic
equilibrium and taking the inner product with three
displacements, we find the following equality
\ba
&&\xi^a \xi^b \xi^c \left( \nabla_a \nabla_b \nabla_c \Phi
	+ \frac{1}{\rho} \nabla_a \nabla_b \nabla_c p \right) = \nonumber \\
&& \qquad
- \xi^a \xi^b \xi^c \left( 
\frac{(\gamma+1)}{\rho\gamma^2p^2} \nabla_a p \nabla_b p \nabla_c p
+ \frac{1}{\rho\gamma^2 p} \nabla_ap \nabla_b p \nabla_c \gamma
- \frac{3}{\gamma\rho p} \nabla_a\nabla_b p \; \nabla_c p \right).
\label{deriveq}
\ea

The hydrostatic equilibrium equation (\ref{deriveq}) and the integration
by parts formula (\ref{triplep}) can now be substituted into the equation
for the third order Hamiltonian (\ref{deltaE}) resulting in the
equivalent third order Hamiltonian
\ba
\delta H_3 
	&=& \frac16 \int dV \left[
		 \; \delta p \;
		\left(  \nabla_a \xi^b \nabla_b \xi^a 
		- \frac{1}{p\gamma} \xi^a \xi^b \nabla_a \nabla_b p \right) \right. \nonumber \\
	&& \qquad
		- \xi^a \xi^b \xi^c \left( 
		\frac{(\gamma+1)}{\rho\gamma^2p^2} \nabla_a p \nabla_b p \nabla_c p
		+ \frac{1}{\rho\gamma^2 p} \nabla_ap \nabla_b p \nabla_c \gamma
		\right) \nonumber \\
	&& 
	\qquad + 2  ( \nabla \cdot \xi) \;
		\left( ( \nabla \cdot \xi) \; \xi\cdot\nabla p
		- p \xi\cdot\nabla( \nabla \cdot \xi) \right) \nonumber \\
	&&  \qquad  -3 \; p \; (\Gamma_1-1) \;  \nabla \cdot \xi \;
		\nabla_a\xi^b \nabla_b\xi^a 
		- \; p \; \left( (\Gamma_1-1)^2 + \frac{\partial\Gamma_1}{\partial \ln \rho}\right) \; 
		(\nabla \cdot \xi)^3  \nonumber \\
	&&   \qquad \left. + 3 \;\rho \; \xi^i \xi^j \nabla_i \nabla_j \delta \Phi
	\right] . \label{H3}
\ea
Equation (\ref{H3}) was derived using only the assumption of
rigid rotation. No assumption about the type of perturbation has been used.

It is now of interest to consider the coupling of r-modes at third order in perturbation
theory. The traditional r-modes are solutions of the linearized Euler equation in the
limit of small rotation rate, $\bar{\Omega} = \Omega\sqrt{R^3/M} \le 1$ and the slow 
rotation approximation will be used for the remainder of this paper. In the slow
rotation approximation, the r-mode displacement satisfies (eg. Friedman \& Morsink 1998)
\be
 \nabla \cdot \xi \sim \delta p \sim   \xi \cdot \nabla p \sim
\xi \cdot \nabla \gamma \sim O(\bar{\Omega}^2),
\ee
while in Appendix B it is shown that the terms 
\be
 \nabla_a \xi^b \nabla_b \xi^a \sim  \xi^a \xi^b \nabla_a \nabla_b p  \sim O(1)
\ee
in the slow rotation expansion.

In the slow rotation approximation, adopting the Cowling approximation
 the third order Hamiltonian coupling 
r-modes together is
\be
\delta H_3 
	= \frac16 \int dV 
		 \; \delta p \;
		\left(  \nabla_a \xi^b \nabla_b \xi^a 
		- \frac{1}{p\gamma} \xi^a \xi^b \nabla_a \nabla_b p \right)  + O(\bar{\Omega}^4).
\label{r3Ham}
\ee
Note that the third order Hamiltonian is $O(\bar{\Omega})^2$ in the slow rotation expansion.
(The term dropped by assuming the Cowling approximation is also $O(\bar{\Omega})^2$.)
This has important consequences, since the energy of the r-modes (the term 
$\delta H_2$  in the perturbation expansion) is also $O(\bar{\Omega})^2$ (Friedman \& Morsink 1998).
If the third order term was first order in the dimensionless angular velocity, then
the third order term would dominate over the second order terms whenever the amplitude
of the r-mode grew larger than the square of the angular velocity. Instead, we see that
the third order term in the Hamiltonian is of the same order (in the small angular velocity limit)
as the second order term. As a result, if the third order terms are responsible for
saturation, the limit on the r-mode amplitude which they set will be independent of 
angular velocity. However, this is strictly only true if no dissipation is present. Since
the dissipative timescales will depend on angular velocity, there will always be 
an angular velocity dependence in the saturation amplitude.

In order to investigate the possible saturation of an unstable r-mode
through couplings with other r-modes at third order, it is necessary to solve for the
r-mode at a high enough order in the slow rotation approximation that the pressure
perturbation is non-zero. For if only the lowest order in angular velocity terms are kept,
the r-mode pressure perturbation will vanish and the third order Hamiltonian will
also vanish \citep{Sch01}. 

\section{Nonlinear Equations of Motion}
\label{eom}

The equations of motion for a general perturbation at any order of perturbation theory
can be found in a straight-forward manner
if the equilibrium star is nonrotating. The 
general procedure is to expand the displacement into a sum over normal mode solutions
\be
\xi = \sum_A c_A(t) \zeta_A(x)
\label{exp}
\ee
where $\zeta_A(x)e^{-i\omega_At}$ is a solution of the linearised 
equations of motion. The expansion coefficients, $c_A(t)$ are found by substituting in the 
expansion (\ref{exp}) into Euler's equation and making use of the orthogonality 
properties of the functions $\zeta_A$. 
This procedure is not as straight-forward when the background
star is rotating. The main complication is that the normal mode solutions are not
orthogonal with respect to the usual inner product,
\be
<\zeta_A,\zeta_B> = \int\; dV \; \rho \zeta_A^* \zeta_B  \neq 0 
	\quad \hbox{for}\quad A \neq B.
\ee
If such a procedure were to be followed, the resulting equations of motion would
be coupled at linear order \citep{Sch01}. The coupling of equations at linear
order can be circumvented by using a phase space expansion of the displacement,
which combines equation (\ref{exp}) with the expansion \citep{Sch01}
\be
\dot{\xi} = \sum_A (-i \omega_A) c_A(t) \zeta_A(x).
\ee

The resulting equations of motion derived by
\citet{Sch01} for the expansion coefficients
(including third order terms in the Hamiltonian) are
\be
\dot{c}_A(t) + i \omega_A c_A(t) = \frac{i\omega_A}{2} 
	\sum_{BC} \frac{ \kappa^*_{ABC}}{\epsilon_A} c_B^*(t) c_C^*(t)
\label{eqmotion}
\ee
where $\epsilon_A$ is the mode's energy in the rotating frame
at unit amplitude defined by
\be
\delta H_2 = \sum_A \epsilon_A |c_A|^2
\ee
and the 
nonlinear coupling coefficient $\kappa$ is defined by
\be
\delta H_3 = - \frac13\sum_{ABC} c_A(t) c_B(t) c_C(t) \kappa_{ABC}.
\ee


There is freedom to choose the amplitudes of the spatial mode functions
$\zeta$. One simple scheme is to choose these amplitudes so that
the rotating frame energies all take the value
\be
\epsilon_A = M R^2 {\Omega}^2 
\ee
which is the same order of magnitude as the star's kinetic energy. 
With this choice, $|c_A|^2$ is the ratio of the mode's energy
to the star's energy. Since the coupling coefficients $\kappa_{ABC}$
have units of energy, the fraction $\kappa_{ABC}/\epsilon_{A}$
corresponds to a dimensionless fraction of the star's energy.
When the mode amplitudes satisfy
\be
\frac{|c_B||c_C|}{|c_A|} \sim \frac{\epsilon_A}{\kappa_{ABC}}
\label{Amplimit}
\ee
the nonlinear terms in the equation of motion (\ref{eqmotion}) will dominate over the
linear terms, which signals the breakdown of the weakly nonlinear regime. 
If (\ref{Amplimit}) is satisfied, it will be necessary to include higher
order nonlinear terms in the equations of motion.

\subsection{Coupling between r-modes}

In the case of the r-modes, the coupling coefficients are
\ba
\kappa_{ABC} &=&  -\frac16\int \; dV \left[
	 \; \delta p_A \;
		\left(  \nabla_a \zeta_B^b \nabla_b \zeta_C^a 
		- \frac{1}{p\gamma} \zeta_B^a \zeta_C^b \nabla_a \nabla_b p \right) \right.  \label{kappa-r}\\
	  && \qquad \left. + \delta p_B \;
		\left(  \nabla_a \zeta_C^b \nabla_b \zeta_A^a 
		- \frac{1}{p\gamma} \zeta_A^a \zeta_C^b \nabla_a \nabla_b p \right) 
	 + \delta p_C \;
		\left(  \nabla_a \zeta_A^b \nabla_b \zeta_B^a 
		- \frac{1}{p\gamma} \zeta_A^a \zeta_B^b \nabla_a \nabla_b p \right) 
	\right]\nonumber
\ea
where $\delta p_A = 	-\Gamma_1 p \nabla\cdot\zeta_A - \zeta_A\cdot\nabla p$.
The coupling coefficients for the case of three r-modes are of second order in 
angular velocity. In the case of r-mode couplings, a dimensionless coupling 
coefficient, $\bar{\kappa}_{ABC}$ is defined by
\be
\bar{\kappa}_{ABC} = \frac{\kappa_{ABC}}{MR^2{\Omega}^2},
\label{dimless}
\ee
so that the magnitude of the $\bar{\kappa}_{ABC}$ coefficients are independent
of the star's angular velocity. This choice of dimensionless coupling
coefficient also has the nice feature that the
coefficients are independent of the star's mass, once a polytropic model
has been chosen.

The r-mode frequencies, given quantum numbers $l_A$ and $m_A$,  are 
\be
\omega_A = \frac{2 m_A \Omega}{l_A(l_A+1)} + C(l_A,m_A,k_A) \Omega\bar{\Omega}^2  ,
\label{r-freq}
\ee
where the frequency correction $C(l_A,m_A,k_A)$ depends on $k_A$, 
the number of radial nodes in the eigenfunction, as well
as the equation of state and buoyancy law.  
It is convenient to introduce 
dimensionless frequencies $\bar{\omega}_A$
are defined by
\be
\bar{\omega}_A = \frac{\omega_A}{\Omega}.
\ee

With the new dimensionless coefficients and frequencies, the equations 
of motion coupling r-modes together are
\be
\dot{c}_A(t) + i \bar{\omega}_A c_A(t) = \gamma_A c_A(t) + 
	\frac{i\bar{\omega}_A}{2} 
	\sum_{BC} \bar{\kappa}^*_{ABC} c_B^*(t) c_C^*(t)
\label{dimeqmot}
\ee
where the time coordinate is now measured in units of the star's spin
period, and an external damping/driving term proportional to $\gamma_A$
has been included. 
When the external damping and driving terms are neglected, 
equation (\ref{dimeqmot}) is independent of angular velocity.
As a result, as the star spins down, the nonlinear coupling equations
are unchanged if no dissipation is present. When dissipation is 
included, the strength of the dissipation terms changes with time 
if the star spins down. 
The scalings with angular velocity which have been 
used are only strictly correct 
in the slow rotation approximation. However, calculations of r-modes
for rapidly rotating stars \citep{LI99,Yos00,Kar00}
have shown that these scalings are not a bad approximation when
the star is rapidly rotating.

\section{Resonance Condition and Selection Rules}
\label{res}

The coupling coefficients $\kappa_{ABC}$ involve integrals of 
three r-mode eigenfunctions over all space. This leads to two 
selection rules on the possible values of $l$ and $m$ for each
mode. Integration over $\phi$ gives the selection rule on
the azimuthal quantum numbers,
\be
m_A + m_B + m_C = 0.
\label{msel}
\ee
Since the integration is over all space, the overall parity of
the integrand must be even. Since r-modes have axial parity,
the parity of an r-mode with angular momentum quantum number 
$l$ is $(-1)^{l+1}$. It follows then that the only allowed
couplings of r-modes must obey
\be
l_A + l_B + l_C + 1 = 0 \quad \hbox{(mod 2)}.
\label{lsel}
\ee

It is simple to see that as a result of these selection rules,
there are no couplings between r-modes of barotropic stars
(ie. neutrally convective stars)
in third order perturbation theory. Stars which are barotropic
only have r-mode solutions with $l=|m|$. The $l$ selection rule
(\ref{lsel}) in this case is
\be
|m_A| + |m_B| + |m_C| + 1 = 0 \quad \hbox{(mod 2)}.
\ee
However it is a property of the integers that 
\be
|m_A| + |m_B| + |m_C| = m_A + m_B + m_C  \quad \hbox{(mod 2)}.
\ee
In order to have nonzero coupling coefficients between
r-modes of barotropic stars, we would require 0 = 1 (mod 2),
which is clearly impossible \citep{Sch01}.

One might expect that the usual triangle inequality which
occurs in the addition of angular momentum should hold here
as well. However, since the formula for the coupling coefficients
involves the pressure perturbation, the triangle inequality is 
modified. If the lowest-order axial term in the r-mode expansion
corresponds to quantum number $l_C$, the pressure perturbation will
have quantum number $l_C \pm 1$ \citep{Sai82}. The modified triangle inequality is then
\be
l_C - 1 \le l_A + l_B  \quad \hbox{and} \quad |l_B-l_A| \le l_C +1.
\ee

In order to find nonzero coupling coefficients between r-modes, 
the perturbations must have a nonzero Schwarzschild discriminant
so that r-modes with $l\neq |m|$ may exist. Including a Schwarzschild
discriminant in the perturbation theory is equivalent to introducing
a source of buoyancy. Some physical mechanisms for the inclusion
of buoyancy in neutron stars which have been suggested are
finite temperature \citep{MVS83}, 
composition gradients \citep{RG92},
and accretion \citep{BC95}.  The 
general effect of buoyancy can be modelled by introducing
an adiabatic index $\Gamma_1$ different from the polytropic 
index $\gamma = 1 + 1/N$. The dimensionless parameter
$1 - \gamma/\Gamma_1$ is typically quite small for all
of these buoyancy models. For instance, for the composition
gradient buoyancy introduced by \citet{RG92}, 
\be
1 - \frac{\gamma}{\Gamma_1} \sim 3 \times 10^{-3} \frac{\rho}{\rho_{nuc}}
\ee
in the core of the star, where $\rho_{nuc} = 2.8\times 10^{14} \hbox{g} \hbox{cm}^{-3}$. 
The computation of r-modes including
a realistic adiabatic index which varies with position
is a challenging  problem, since the equations for the 
$l \neq |m|$ r-modes are not valid at any places where
$1 - \gamma/\Gamma_1$ vanishes. At present, no calculation
of r-modes of neutron stars with varying $\Gamma_1$
 exists in the literature. A reasonable 
compromise is a simpler model in which $\Gamma_1$ is
a constant. The r-modes of polytropic stars 
with constant $\Gamma_1$ have been examined in great
detail by \citet{YL00}. Following the work of 
\citet{YL00}, we will also adopt similar buoyancy
models in the present paper. An important property 
of the $l \neq |m|$ r-modes has been pointed out
in the work of \citet{YL00}: the $l \neq |m|$ r-modes
only exist in the limits of slow rotation 
compared to the break-up velocity ($\bar{\Omega}\ll1$)
and slow rotation compared to the Brunt-Vaisala
frequency  which places an approximate
upper limit of $\bar{\Omega} \sim \sqrt{1 - \gamma/\Gamma_1}$
on the angular velocity.

In order to make fully nonlinear numerical simulations 
computationally possible, the physics is generally 
simplified to the level of a perfect, zero-temperature
fluid, as in the simulations by \citet{SF01} and 
\citet{LTV01a,LTV01b}. For this reason, these simulations can not include
any source of buoyancy and the stars can only support the oscillations
of r-modes with $l=|m|$. Since the arguments presented above show
that there are no nonlinear couplings between $l=|m|$ r-modes
at third order in perturbation theory, the fully nonlinear codes 
do not access the same physical coupling mechanisms explored in
the present paper. At linear level, the codes written by
\citet{SF01} and \citet{LTV01a,LTV01b} model the perturbations 
known as hybrid modes \citep{LF99}. Nonlinear couplings between the
hybrid modes have been computed by \citet{Sch01}. The hybrid
modes are good description of a star's low-frequency 
modes when the Brunt-Vaisala frequency is very small compared to
the spin frequency.

It is well known
from the theory of nonlinear oscillations that whenever a 
resonant match of frequencies occurs in a system consisting
of three coupled oscillators, it is possible to transfer 
energy from a large amplitude mode to a small amplitude mode.
The condition for resonance is that the detuning, defined by
\be
\Delta \omega_{ABC} = \omega_A + \omega_B + \omega_C ,
\ee
should be close to zero.
If any resonances occur amongst the
direct couplings with the $l=m=2$ r-mode, the unstable mode's 
amplitude could be limited by the coupling, even if the
coupling coefficient is small.

In the numerical simulations of \citet{SF01} and \citet{LTV01a,LTV01b},
the initial amplitude of the unstable $l=m=2$ r-mode is set to a small
value and all other mode amplitudes are set to zero. In terms of 
the equations of weakly nonlinear system, this corresponds to the
$l=m=2$ r-mode acting a a ``parent'' or source for a later generation
of modes. As an example, the parent mode will carry the label $A=0$. Since all other modes 
initially have zero amplitude, the first generation of daughter modes
will be only those with nonzero coupling coefficients $\kappa_{100}$,
where the label $1$ refers to an excited daughter mode. In the case 
of the CFS instability of r-modes, the unstable parent mode
is the $l=|m|=2$ r-mode. The daughter modes which can be excited
through the third order coupling can be found using the selection rules
(\ref{msel}) and (\ref{lsel}), choosing $l_A=l_B=2$, $m_A=m_B=2$ and
solving for $l_C$ and $m_C$. The daughter modes which can be
excited are shown in Table \ref{table1}. 
The detuning can be easily calculated by using the formula
for the r-mode frequency in the slow rotation limit $\bar{\Omega}\ll 1$.
For this 
first generation of couplings, with $l_A=m_A=l_B=m_B=2$, and
$m_C=-4$, the detuning is
\be
\frac{\Delta \omega_{ABC}}{\Omega} = 
\frac43 -\frac{8}{l_C(l_C+1)} = \frac{4(l_C+3)(l_C-2)}{3l_C(l_C+1)},
\ee
which is non-zero, since the selection rules require $l_C = 5$.

Similarly, a second generation of daughter modes can be excited
through couplings of first generation daughters with the unstable
r-mode or through the coupling of two daughter modes.
The second generation daughter modes are also displayed in Table \ref{table1}.
It should be clear that modes excited at any generation must have an 
even value of $m$. Furthermore, it is impossible to generate certain modes,
such as an r-mode with $m=2$ and an odd value of $l$ if the only nonzero
initial amplitude corresponds to the unstable $l=m=2$ r-mode.
This type of coupling, where the parent mode acts as a source for 
a daughter mode will be denoted a direct coupling. 
In the case of
the second generation direct couplings, the detuning is either
\be
\frac{\Delta \omega_{ABC}}{\Omega} = 
\frac{2 \left(l^2 + l + 10\right)}{5l(l+1)} > 0
\ee
in the case of couplings involving $l_B=5$ and $m_B=-4$ 
or
\be
\frac{\Delta \omega_{ABC}}{\Omega} = 
\frac{2 \left(7l^2 + 7l -6\right)}{15l(l+1)} 
\ee
which is also nonzero for whole number values of $l$.
This shows that a numerical simulation which sets to 
zero the initial values of all modes except for the
$l=m=2$ r-mode will only involve non-resonant interactions.

\begin{figure}
\figurenum{1}
\label{fig1}
\epsscale{0.8}
\plotone{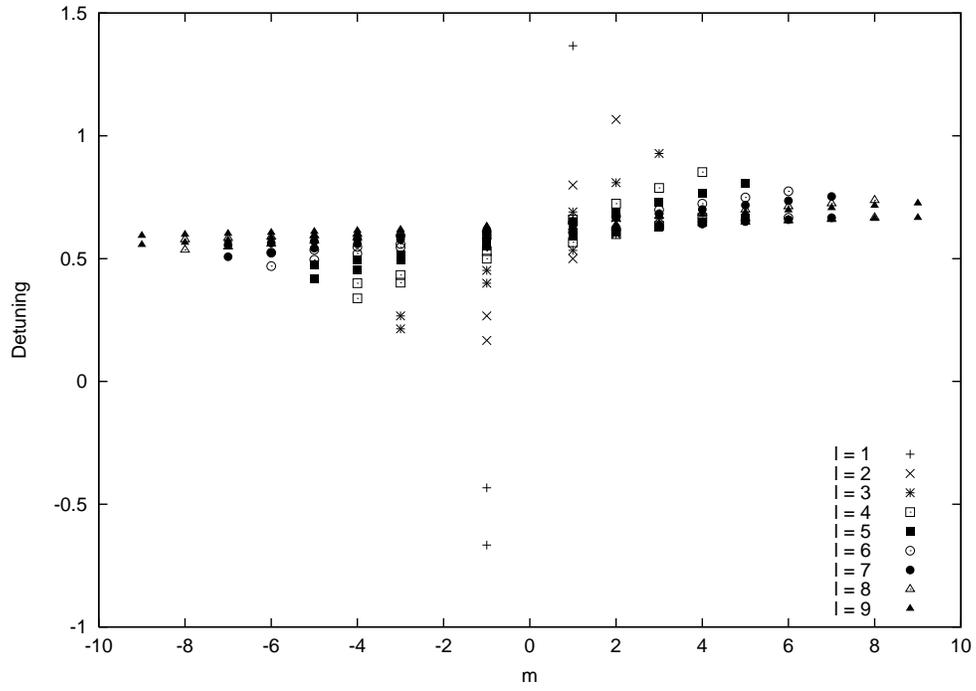}
\caption{Plot of the detuning versus $m$ for values of $l\le10$. The
detuning is defined in equation (\ref{detune}).}
\end{figure}

In a realistic star it is expected that the other modes will have small
but nonzero amplitudes. When is 
it possible to have a resonance in any triplet of r-modes including the
$l_A=m_A=2$ r-mode? Suppose we choose values $l_C=l$ and $m_C=m$  for the quantum numbers
for mode C. From the m-selection rule $m_B = -(m+2)$ and if we
choose $l_B\ge l$ the l-selection rule and the triangle inequality
restrict $l_B$ to the form $l_B = l + 2k + 1$ where $k$ is either 0 or
1. The general formula for the detuning in this case is
\be
\frac{\Delta \omega_{ABC}}{\Omega} = 
\frac23 - \frac{4}{l_B(l_B+1)}
	+ \frac{4 m (2k+1)(l+k+1)}{l(l+1)l_B(l_B+1)}.
\label{detune}
\ee
In Figure \ref{fig1} it can be seen that 
the detuning can't vanish if 
we are restricted to nonzero values of $m$ and $m_B$. 
Hence there are no direct resonant couplings with the
$l=m=2$ r-mode. However, the values of the detuning
are typically smaller than unity and 
quickly converge to $2/3$ for large values of $l$.

Any three r-modes with an identical odd value of $l$ will be resonant
if they satisfy the m-selection rule. This occurs because the 
detuning for modes with $l_A=l_B=l_C=l$ is
\be
\frac{\Delta \omega_{ABC}}{\Omega} = 
\frac{2}{l(l+1)}\left(m_A + m_B + m_C\right)
\ee
which vanishes when equation (\ref{msel}) is satisfied. The 
$l$-selection  rule (\ref{lsel}) demands that $l$ must be odd,
so that these resonant triplets do no involve the $l=m=2$ mode.
Since the resonance criteria occurs for all odd values of $l$,
there are an infinite number of resonances, none of which
couple directly with the $l=m=2$ r-mode. These couplings 
are indirect resonant couplings. These resonances will 
not be exact, since the r-mode frequencies have the small
correction terms shown in equation (\ref{r-freq}). In the
slow rotation limit, however, these corrections are very
small, and these triads will be very close to resonance.

\section{Numerical Results}
\label{num}

In this paper, the r-mode eigenfunctions of polytropic stars are 
solved by a numerical method, similar to the 
method described by \citet{Sai82}, which keeps terms in the
perturbed velocity that are third order
in the angular velocity. The code's accuracy was
checked against the results published by \citet{Sai82} for the
case of modes where $\gamma \neq \Gamma_1$ and against the
results of \citet{LMO99} for the case of modes of stars with
$\gamma = \Gamma_1$. Good agreement was found in both cases. 
Once the r-mode eigenfunctions for large number of modes has
been found, the coupling coefficients can be computed by
directly integrating equation (\ref{kappa-r}). In these calculations,
all r-modes with $l\le 10$  and  less than 3 radial nodes were computed.

There are two types of three-mode couplings which occur amongst
the r-modes of buoyant stars. Triplets of modes which include
the $l=m=2$ r-mode are always nonresonant. These direct nonresonant
mode couplings will be examined in section \ref{dirnonres}.
Resonant mode couplings can occur triplets which do not include the
$l=m=2$ r-mode. This type of coupling is an indirect resonant 
coupling and will be examined in section \ref{indirres}.

\subsection{Direct Nonresonant Mode Couplings}\label{dirnonres}

Table 1 shows the dimensionless coupling coefficients for generating the first
and second generations of r-modes. The equilibrium stellar model is a $N=1$ 
polytrope and the adiabatic index $\Gamma_1 = 1.9$. The $N=1$ polytrope was
chosen to allow the closest comparison with the nonlinear simulations
 of \citet{SF01,LTV01a,LTV01b}
who all make use of the same polytropic model. However, since  $\Gamma_1 \neq
1 + 1/N$, the types of couplings discussed in this paper will not be seen in the
nonlinear simulations. The size of the coupling coefficients in Table 1 are all
small. 
Equations (\ref{Amplimit}) and (\ref{dimless}) 
imply that the unstable r-mode would have to 
grow to an amplitude so that the fraction of energy in the mode to the star's 
energy is of order $1/\bar{\kappa}^2$ before higher order nonlinear terms 
would have to be considered. Since the coupling coefficients are of
order $10^{-2}$, this suggests that the unstable r-mode would need to have 
an energy which is approximately $10^4$ times the energy of the star
before higher order nonlinear terms become important. In other words,
the weakly nonlinear approximation holds for all physical values of mode
energy.
 
The dependence of the coupling coefficients on the equation of state 
and buoyancy law was also examined. The coupling coefficients corresponding to
the first generation mode coupling (for modes with zero radial nodes) are
shown in Table \ref{table2}. There appears to be very little dependence on
the equation of state or on buoyancy. The lack of dependence on buoyancy 
can be understood by considering  equation (\ref{kappa-r}) which defines
the coupling coefficients. The coupling coefficients only depend on 
equilibrium quantities (the pressure) and the eigenfunctions of the perturbations.
There is no explicit dependence in (\ref{kappa-r}) on the eigenvalue $C(l,m,k)$ 
defined in equation (\ref{r-freq}). As shown by \citet{YL00}, the eigenvalues and
eigenfunctions of the $l=|m|$ r-modes have very little dependence on the 
buoyancy law. The only dependence on buoyancy can come through the $l>|m|$ 
r-modes. For the eigenvalues of these modes there is a strong dependence 
on buoyancy, which we have found varies approximately as 
$C(l,m,k) \sim (\Gamma_1 - \gamma)^{-1}$. (This scaling can be also
be deduced from the form of the equations presented in the work of \citet{Sai82}.)
 However, as long as the eigenvalue
$C(l,m,k)$ and the dimensionless angular velocity $\bar{\Omega}$ are small enough that 
$C(l,m,k) \bar{\Omega}^2 << 2m/l(l+1)$, the form of the eigenfunctions are almost 
independent of the buoyancy. In order to satisfy this inequality for angular velocities
as large as $\bar{\Omega} \sim 0.1$, we require that $\Gamma_1-\gamma > 10^{-3}$.
 As the difference $\Gamma_1 - \gamma$
decreases to zero, the $l>|m|$ r-modes cease to exist \citep{YL00} so we would expect that
at small values of $\Gamma_1 - \gamma$ the character of these r-modes will be greatly 
modified and the values of the coupling coefficients will be altered. In Table~2
the values of the coupling coefficients for $\gamma=2$ are practically the same for
all values of $\Gamma_1$. It is only once $\Gamma_1 - \gamma$ falls below $8\times10^{-4}$
that any variation in the coupling coeffients is found.

The largest coupling coefficients for modes which couple to the $l_A=m_A=2$ r-mode
are shown in Table {\ref{table3}}. The largest coefficients appear to be those
which couple $(l_B,m_B) = (l,-1)$, with $(l_C,m_C) = (l+1,-1)$. 
 As long as these modes have a small
but nonzero amplitude, it will be possible for these modes to be excited. 
In this case, these modes could be the most important type of interaction
which could conceivably limit the amplitude of the unstable r-mode.

As with all other types of modes of nonbarotropic stars, there are an infinite
number of r-mode solutions for each value of $l$ and $m$. Each solution 
can be labelled by the quantum number $k \ge 0$ which corresponds to the
number of radial nodes in the eigenfunction. The CFS instability in the
r-modes is strongest in the fundamental $k=0$ modes, so only the fundamental
$l=m=2$ r-mode will be considered in this work. However, it is conceivable that
the unstable mode could excite daughter modes with $k>0$. In Table 1, the 
dependence on number of nodes for the first generation daughter mode is shown. 
It is intriguing that the magnitude of the coefficients does not seem to fall
off rapidly with increasing node number. This behaviour is due to the weighting
of the integrals in the coupling coefficients, which favours the regions of
the eigenfunctions closest to the surface of the star. Since most of the
nodes in a $k>0$ eigenfunction occur in the middle regions of the star,
and the eigenfunctions are peaked near the surface, there is not a large 
difference in the contributions of different $k$-eigenfunctions to the
coupling coefficient integrals. All of these higher-order eigenfunctions 
are able to participate in interactions with the $l=m=2$ r-mode. This 
increases the number of modes which can act as a drain of 
the unstable mode's energy. In order to evaluate an 
upper bound on $N$, it is necessary 
to compute high-radial order r-modes, which 
becomes difficult since these modes oscillate 
rapidly. A more suitable approach would be an approximation
scheme such as the WJKB method. This type of approach would
allow an asymptotic calculation of the coupling coefficients
for high-radial order modes. However, it seems unlikely that
the high-radial order modes would participate in the type of 
interactions described above, for they would be rapidly 
damped by viscosity. In this case of a series of modes
with approximately equal frequencies and coupling coefficients,
the most slowly damped mode is the one which places the
strongest limit on the unstable parent mode \citep{WG01}. 
In this case, the fundamental mode is damped the least 
by viscosity, so it should be the most important.

\subsection{Indirect Resonant Couplings} \label{indirres}

In section \ref{res} it was shown that there are an infinite
number of resonant triads which do not involve the 
$l=m=2$ r-mode. These resonant triads only couple indirectly
to the unstable r-mode. 

The coupling coefficients for resonant mode interactions 
are shown in Table \ref{table4}. We note that the 
coupling coefficients in Table \ref{table4} are at least an order of
magnitude larger than the coefficients in Table \ref{table1}. Due
to the infinite number of possible couplings of this type, only
a small selection of triplets are shown in Table \ref{table4}.

In order to examine the possible effect of indirect
resonant couplings on the amplitude of the $l=m=2$
r-mode, we can consider a toy model consisting of only
three modes. Mode A has quantum numbers $l_A=m_A=2$.
Mode B has quantum numbers $l_B=5$ and $m_B=-4$. Modes
A and B can interact through a triplet involving two
of mode A and one of mode B, with coupling coefficient
$k_A := \kappa_{AAB}$. The AAB triplet is a nonresonant 
interaction, for the reasons discussed in the previous
subsection. The third mode, mode C, has quantum
numbers $l_C = 5$ and $l_C = 2$. Mode C can interact
with mode B through a triplet involving two of mode C
and one of mode B, with coupling coefficient 
$k_C := \kappa_{CCB}$. The CCB coupling is resonant. 
Interactions between A and C are not allowed by
the selection rules discussed at the beginning of
this section.

Since the CCB coupling is resonant, it is 
possible for an efficient transfer of energy
to occur which would limit the amplitude of mode B. 
Suppose that  mode B is the most important mode for saturating
mode A when only triplets interacting with A are considered. 
As soon as mode C is included, the rapid transfer
of energy between B and C could drastically alter
the saturation of mode A. This type of situation is 
generic, since all allowed triplets involving the
$l=m=2$ r-mode also involve one r-mode with an odd
value of $l$. Since r-modes with odd $l$ always have 
resonances (except for $l=1$), this type of indirect
resonant interaction is possible.
A similar 
situation was encountered in the study of 
g-modes by \citet{WG01}. In their study, \citet{WG01}
found that such couplings slowed down the transfer
of energy between modes A and B, but were unable
to halt the energy transfer. A similar study for
r-modes will be the topic of a future investigation.

\section{Conclusions}
\label{conclusions}

In this paper, we derived an alternative form of the Hamiltonian describing the
interaction of r-modes at lowest nonlinear order. This Hamiltonian was used
to compute coupling coefficients between r-modes of polytropic stars with
a source of buoyancy. In the case that only the $l=m=2$ CFS unstable r-mode
has a nonzero amplitude, none of the couplings are resonant, and the
coupling coefficients are all small. If small but nonzero amplitudes are
allowed for the damped r-modes, larger couplings can occur. These
triads of modes with the largest coupling coefficients are the ones most
likely to contribute to the saturation of the unstable r-mode's amplitude.
An unusual aspect of this problem which we have pointed out is that there
are an infinite number of resonant triads which couple indirectly to
the unstable r-mode. It should be noted that all of the couplings considered
in this paper involve the $l>|m|$ r-modes which only exist in the limit
of slow rotation. This study is complementary to the work of \cite{Sch01} who examined the
coupling coefficients for the modes of rotating stars which are neutral
to convection, which is a very good approximation when the star is rapidly rotating.
The results of a numerical evolution of the coupled system of oscillators considered in
this paper and in the paper by \citet{Sch01} 
will be presented in a future paper \citep{Arr01}. 

This paper has been concerned only with the couplings between different
r-modes. However, it may be possible to couple r-modes with other modes,
such as the g-modes or p-modes. The g-modes are highly distorted by
rotation \citep{BUC96}, so the effect of their coupling with r-modes 
is difficult to compute in the slow-rotation approximation. 
Instead, for rapid rotation, the inertial
or hybrid modes discussed by \citet{LF99}, and \cite{YL00} are a better approximation
to the g-modes and r-modes. Hence the coupling between hybrid modes
studied by \citet{Sch01} should be sufficient to study all of the
different types of couplings between different families of modes. 

When the driving and damping of modes is included in the evolution
equations describing the coupled oscillators, it is possible for 
equilibrium solutions with constant amplitude to result 
\citep{Dzi82,DK84,WG01}. This leads to a limiting amplitude 
on unstable modes. This limiting amplitude depends on the coupling
coefficients, frequency detuning and the driving and damping rates
of the modes. With the state of understanding of these 
parameters coming from this paper and the work of \citet{Sch01},
it will be possible for numerical simulations of mode-mode
coupling to show under what conditions it is possible for 
saturation of the unstable r-mode to occur.

\acknowledgments

It is a pleasure to thank Phil Arras, 
\'{E}anna Flanagan, John Friedman, Keith Lockitch,
Katrin Schenk, Saul Teukolsky and Ira Wasserman
for many helpful discussions. I would like to thank the Institute
for Theoretical Physics for hospitality during the workshop
``Spin and Magnetism in Young Neutron Stars'' and
the Canadian Institute for Theoretical Astrophysics where
some of this research was carried out. 
This research was supported by the Natural Sciences and Engineering
Research Council of Canada and by NSF grants PHY99-07949
and PHY95-07740.

\appendix
\section{Integration by Parts}
\label{parts}

In equation (\ref{deltaE}) the integral
\be
I = \int dV p 
	\nabla_a \xi^b \nabla_b \xi^c \nabla_c \xi^a
\label{I}
\ee
appears. The purpose of this appendix is to transform the
integral (\ref{I}) into a form more suitable for 
evaluating the r-mode coupling coefficients.
The goal is to transfer the derivatives acting on the 
displacements to derivatives acting on the pressure. This
can be done by integrating by parts a many times.
All surface terms are  proportional to either
the pressure or the radial derivative of the pressure, both
of which vanish on the surface of the star.

Integrating by parts, and throwing away the surface terms,
\be
I = - \int dV \xi^a \nabla_c 
	\left( p \nabla_a\xi^b \nabla_b \xi^c \right) 
	= I_A + I_B + I_C,
\ee
where
\ba
I_A &=& - \int dV \nabla_c p \;\xi^a \nabla_a\xi^b \; \nabla_b \xi^c \\
I_B &=& - \int dV p \;\xi^a \nabla_a\nabla_c \xi^b \; \nabla_b \xi^c \\
I_C &=& - \int dV p \xi^a \nabla_a \xi^b \; \nabla_b(\nabla\cdot\xi).
\ea

The integral $I_A$ has one derivative of pressure appearing. 
Integrating by parts once more will produce a term involving 
two derivatives of pressure, since
\be
I_A = \int dV \xi^c \nabla_b
	\left( \nabla_c p  \xi^a \nabla_a \xi^b \right)
	= I_{A1} + I_{A2} + I_{A3},
\ee
where 
\ba
I_{A1} &=& \int dV \xi^c \nabla_c \nabla_b p \; 
	\xi^a \nabla_a \xi^b\\
I_{A2} &=& \int dV \xi^c \nabla_c p \; 
	\nabla_a \xi^b \nabla_b \xi^a\\
I_{A3} &=& \int dV \xi^c \nabla_c p \;
	\xi^a \nabla_a( \nabla \cdot \xi ).
\ea

The integral $I_{A1}$ now has two derivatives acting on the pressure.
Only one more integration by parts will be necessary to provide the
term involving three covariant derivatives of pressure.
The last integration by parts gives 
\ba
I_{A1} &=& - \int dV  \xi^b \nabla_a 
	\left( \xi^a \xi^c \nabla_c \nabla_b p \right) \\
&=& - \int dV \left(  \nabla \cdot \xi \; \xi^b \xi^c \nabla_c \nabla_b p  
+ \xi^a \nabla_a \xi^c \; \xi^b \nabla_b \nabla_c p
+ \xi^a \xi^b \xi^c \nabla_a \nabla_b \nabla_c p \right).
\ea
Notice that the second term is just the negative of the integral
$I_{A1}$. This allows us to rearrange the last equation to read
\ba
I_{A1} &=& -\frac12 \int dV 
	\left( (\nabla \cdot \xi) \; \xi^b \xi^c \nabla_c \nabla_b p
	+ \xi^a \xi^b \xi^c \nabla_a \nabla_b \nabla_c p \right) .
\ea
This step gives us the term involving three derivatives 
of the pressure. 

The integral $I_B$ involves two derivatives of the displacement, which can 
be simplified through an integration by parts to 
\ba
I_B &=& \int dV  \nabla_c \xi^b \nabla_a 
	\left( p \xi^a \nabla_b \xi^c \right) \\
&=& \int dV \left(\nabla \cdot (p\xi) \nabla_a \xi^b \nabla_b \xi^a 
+ p \nabla_c \xi^b 
	\xi^a \nabla_a \nabla_b \xi^c\right).
\ea
The last term in this expression is just the negative of $I_B$. 
Hence
\be
I_B = \frac12 \int dV \nabla \cdot (p\xi) \nabla_a \xi^b \nabla_b \xi^a.
\ee

The final task is to transform $I_C$ into a 
more useful form. This will be done by integrating by
parts again, giving
\be
I_C = \int dV \nabla \cdot \xi \nabla_b
	\left( p \xi^a \nabla_a \xi^b \right)
	=  I_{C1} + I_{C2} + I_{C3},
\ee
where 
\ba
I_{C1} &=& \int dV \nabla \cdot \xi\; \nabla_b p  \xi^a \nabla_a \xi^b\\
I_{C2} &=& \int dV p\;\nabla \cdot \xi\; \xi^a \nabla_a (\nabla \cdot \xi)\\
I_{C3} &=& \int dV p\;\nabla \cdot \xi\; \nabla_a \xi^b \nabla_b \xi^a.
\ea

The final integration by parts allows us to rewrite $I_{C1}$ 
in the form
\ba
I_{C1} &=& - \int dV \xi^b \nabla_a
	\left( (\nabla\cdot\xi) \xi^a \nabla_b p \right) \\
&=& - \int dV \left( \xi\cdot\nabla p \;\xi\cdot\nabla(\nabla\cdot\xi)
+ (\nabla\cdot\xi)^2\; \xi \cdot \nabla p 
+ (\nabla\cdot\xi)\; \xi^a \xi^b \nabla_a\nabla_b p\right).
\ea

Putting together the results for $I_A$, $I_B$ and $I_C$, the original
integral is
\ba
I &=& -\frac12 \int dV \left( \xi^a \xi^b \xi^c \nabla_a \nabla_b \nabla_c p
	+ 3 (\nabla \cdot \xi) \; \xi^b \xi^c \nabla_c \nabla_b p
	- 3 \nabla \cdot (p\xi) \nabla_a \xi^b \nabla_b \xi^a  \right) \nonumber\\
&& \qquad - \int dV \left( (\nabla\cdot\xi)^2\; \xi \cdot \nabla p 
- p\;(\nabla \cdot \xi) \; \xi^a \nabla_a (\nabla \cdot \xi) \right)
\label{finalI}
\ea
While equation (\ref{finalI}) may not appear to be simpler than equation (\ref{I}),
it will prove to be useful in simplifying the third-order Hamiltonian coupling
r-modes.

\section{Order of Magnitudes for r-modes}

In the slow-rotation approximation, defined by $\bar{\Omega} \ll 1$,
the fluid displacement vector for r-modes has components which are
of order
\be
\xi^r = O(\bar{\Omega}^2) \;, \xi^\theta = O(1) \;, \xi^\phi = O(1).
\ee
In the slow rotation approximation, the star's pressure deviates
from spherical symmetry at second order in angular velocity,
so that derivatives of the pressure are of order
\be
\frac{\partial}{\partial r} p = O(1) \;,
\frac{\partial}{\partial \theta} p = O(\bar{\Omega}^2) \;,
\frac{\partial}{\partial \phi} p = 0 .
\ee
It then follows that 
\be
\xi^a \nabla_a p = \xi^a \partial_a p = O(\bar{\Omega}^2)
\ee
for r-modes.

In equation (\ref{r3Ham}) for the third-order r-mode Hamiltonian,
a term involving two covariant derivatives of pressure appears. In
terms of partial derivatives and Christoffel symbols for spherical
coordinates, this term is
\be
\xi^a \xi^b \nabla_a \nabla_b p 
	= \xi^a \xi^b ( \partial_a \partial_b p - \Gamma^c_{ab} 
	\partial_c p).
\label{d2p}
\ee
The non-vanishing Christoffel symbols for flat space written in
spherical coordinates are
\be
r^2 \Gamma^{\theta}_{r\theta} = 
r^2 \Gamma^{\phi}_{r\phi} = 
- \Gamma^{r}_{\theta\theta} = 
- \frac{1}{\sin^2\theta} \Gamma^{r}_{\phi\phi} = r,  \qquad
\Gamma^{\phi}_{\phi\theta} = \cot\theta.
\label{Christoffel}
\ee
The leading order behaviour of equation (\ref{d2p}) is then
\be
\xi^a \xi^b \nabla_a \nabla_b p = \frac{1}{r} \partial_r p \left(
\xi^\theta \xi_\theta + \xi^\phi \xi_\phi \right) 
 = O(1).
\ee
Similarly, a longer but straight-forward calculation shows that
\be
\nabla_a \xi^b \nabla_b \xi^a = O(1).
\ee

\begin{deluxetable}{cccccccccccccc}
\tablewidth{0pc}
\tablecolumns{14}
\tablecaption{Non-resonant Direct Couplings with an Unstable $l=|m|=2$ r-mode.
The stellar model is a $N=1$ polytrope with $\Gamma_1 = 1.9$. In these tables,
$l$ and $m$ refer to spherical harmonic angular quantum numbers and $k$ refers
the number of radial nodes in the eigenfunction.
\label{table1}}
\tablehead{
\multicolumn{3}{c}{Mode A}&\colhead{} &\multicolumn{3}{c}{Mode B} 
&\colhead{}& \multicolumn{3}{c}{Mode C}   \\
\cline{1-3} \cline{5-7} \cline{9-11}\\
\colhead{l}&\colhead{m}&\colhead{k}&\colhead{}&
\colhead{l}&\colhead{m}&\colhead{k}&\colhead{}&
\colhead{l}&\colhead{m}&\colhead{k}&\colhead{}&\colhead{$\bar{\kappa}_{ABC}$}
&\colhead{$\Delta \omega_{ABC}/\Omega$}
}
\startdata
\cutinhead{First Generation Couplings}
2&      2&      0&&     2&      2&      0&&     5&      -4&     0& &     2.01e-02&1.07\\    
2&      2&      0&&     2&      2&      0&&     5&      -4&     1& &     1.54e-02&1.07  \\ 
2&      2&      0&&     2&      2&      0&&     5&      -4&     2& &     1.33e-02&1.07\\    
2&      2&      0&&     2&      2&      0&&     5&      -4&     3& &     1.20e-02&1.07\\   
2&      2&      0&&     2&      2&      0&&     5&      -4&     4& &     1.11e-02&1.07\\    
2&      2&      0&&     2&      2&      0&&     5&      -4&     5& &     1.04e-02&1.07\\   
2&      2&      0&&     2&      2&      0&&     5&      -4&     6& &     9.69e-03&1.07\\    
2&      2&      0&&     2&      2&      0&&     5&      -4&     7& &     9.58e-03&1.07\\
2&      2&      0&&     2&      2&      0&&     5&      -4&     8& &     9.25e-03&1.07\\
2&      2&      0&&     2&      2&      0&&     5&      -4&     9& &     9.06e-03&1.07\\ 
         \cutinhead{Second Generation Couplings}
2 & 2 & 0 && 5 & -4 & 0 && 4 & 2 & 0 && -8.10e-02 & 0.60 \\
2 & 2 & 0 && 5 & -4 & 0 && 6 & 2 & 0 && 2.33e-01  & 0.49\\
2 & 2 & 0 && 5 & -4 & 0 && 8 & 2 & 0 && 9.83e-04  & 0.45\\
2 & 2 & 0 && 5 & 4 & 0 && 6 & -6 & 0 && -9.68e-02 & 0.91\\
2 & 2 & 0 && 5 & 4 & 0 && 8 & -6 & 0 && 3.05e-02  & 0.92\\
\enddata
\end{deluxetable}

\begin{deluxetable}{ccccc}
\tablewidth{0pc}
\tablecolumns{5}
\tablecaption{Comparison of coupling coefficients 
between two $l=m=2$ r-modes and the $l=5$, $m=-4$ r-mode
for different
polytropic equations of state. \label{table2}} 
\tablehead{
N & $\gamma = 1 + 1/N$ & $\Gamma_1$ & $ 1 - \gamma/\Gamma_1$ & $\bar{\kappa}$ 
}
\startdata
3 & 4/3 & 1.2 & -1/9 &   -3.00e-03	\\
2 & 3/2 & 1   & -1/2 &  -1.21e-02	\\
1 & 2   & 1.99& -5e-03 & 2.16e-02	\\
1 & 2   & 1.9 & -0.05&	 2.01e-02  	\\
1 & 2	& 2.1 &  0.05&   2.35e-02	\\
1 & 2	& 2.01&  5e-03&	 2.20e-02	\\
1 & 2	& 2.0008&4e-04&  4.00e-02 
\enddata
\end{deluxetable}

\begin{deluxetable}{cccccccccccccc}
\tablewidth{0pc}
\tablecolumns{14}
\tablecaption{Largest Coupling Coefficients.
The stellar model is a $N=1$ polytrope with $\Gamma_1 = 1.9$.
\label{table3}}
\tablehead{
\multicolumn{3}{c}{Mode A}&\colhead{} &\multicolumn{3}{c}{Mode B} 
&\colhead{}& \multicolumn{3}{c}{Mode C}   \\
\cline{1-3} \cline{5-7} \cline{9-11}\\
\colhead{l}&\colhead{m}&\colhead{k}&\colhead{}&
\colhead{l}&\colhead{m}&\colhead{k}&\colhead{}&
\colhead{l}&\colhead{m}&\colhead{k}&\colhead{}&\colhead{$\bar{\kappa}_{ABC}$}
&\colhead{$\Delta \omega_{ABC}/\Omega$}}
\startdata 
2 & 2 & 0 && 8 & -1 & 0 && 9 & -1 & 0 && 1.21e+01 &0.62\\
2 & 2 & 0 && 8 & -1 & 1 && 9 & -1 & 1 && 1.07e+01 &0.62\\
2 & 2 & 0 && 9 & -1 & 0 && 10 & -1 & 0 && 1.93e+01&0.63\\
2 & 2 & 0 && 9 & -1 & 1 && 10 & -1 & 1 && 1.71e+01&0.63\\
2&      2&      0&&     10&     -1&     0&&     11&     -1&     0&&      2.92e+01& 0.63
\enddata
\end{deluxetable}

\begin{deluxetable}{ccccccccccccc}
\tablewidth{0pc}
\tablecolumns{13}
\tablecaption{Indirect Resonant Couplings.
The stellar model is a $N=1$ polytrope with $\Gamma_1 = 1.9$.
\label{table4}}
\tablehead{
\multicolumn{3}{c}{Mode A}&\colhead{} &\multicolumn{3}{c}{Mode B} 
&\colhead{}& \multicolumn{3}{c}{Mode C}   \\
\cline{1-3} \cline{5-7} \cline{9-11}\\
\colhead{l}&\colhead{m}&\colhead{k}&\colhead{}&
\colhead{l}&\colhead{m}&\colhead{k}&\colhead{}&
\colhead{l}&\colhead{m}&\colhead{k}&\colhead{}&\colhead{$\bar{\kappa}_{ABC}$}
}
\startdata
5 & -4 & 0 && 5 & 2 & 0 && 5 & 2 & 0 && -1.35e-02\\ 
5 & -4 & 0 && 5 & 2 & 0 && 5 & 2 & 1 && -1.46e-01 \\
5 & -2 & 1 && 5 & 1 & 0 && 5 & 1 & 0 && 1.40e+00  \\
7 & -2 & 0 && 7 & 1 & 0 && 7 & 1 & 1 && -2.08e+00 \\
9 & -2 & 1 && 9 & 1 & 0 && 9 & 1 & 0 && 1.03e+01  
\enddata
\end{deluxetable}

\end{document}